\documentclass[a4paper,12pt]{iopart}

\usepackage{hyperref}
\usepackage{graphicx}
\usepackage{lineno,hyperref}
\usepackage{amssymb}
\usepackage{textcomp}
\usepackage{float}
\usepackage{xcolor}

\begin{document}
	
	\title{New physics from TOTEM's recent measurements of elastic and total cross sections}
	
	\author{Istv\'an Szanyi$^{1,2}$, Norbert Bence$^2$, L\'aszl\'o Jenkovszky$^3$}
	\address{$^1$E\"otv\"os Lor\'and University,\\1/A, P\'azm\'any P\'eter s\'et\'any, Budapest, 1117, Hungary}
	\address{$^{2}$Uzhgorod National University, \\14, Universytets'ka str., Uzhgorod, 88000, Ukraine}
	\address{$^3$Bogolyubov Institute for Theoretical Physics (BITP),\\
		Ukrainian National Academy of Sciences \\14-b, Metrologicheskaya str.,
		Kiev, 03680, Ukraine}
	
	\ead{sz.istvan03@gmail.com, bencenorbert007@gmail.com, jenk@bitp.kiev.ua }
	
	\vspace{10pt}
	\begin{indented}
		\item[]September 2018
	\end{indented}

\newcommand{\keywords}[1]{Keywords: #1}
	
	\begin{abstract}
		We analyze the recently discovered phenomena in elastic proton-proton scattering at the LHC, challenging the standard Regge-pole theory: the low-$|t|$ "break" (departure from the exponential behavior of the diffraction cone), the accelerating rise with energy of the forward slope $B(s,t=0)$, the absence of secondary dips and bumps on the cone and the role of the odderon in the forward phase of the amplitude, $\rho(13~TeV)=0.1\pm0.01$, and especially its contribution at the dip region, measured recently by TOTEM. Relative contributions from different components to the scattering amplitude are evaluated from the fitted model.
		\\
		\\
	\end{abstract}

	\section{Introduction} \label{s1}
	During the past seven  years the TOTEM Collaboration produced a number of spectacular results on proton-proton elastic and total cross sections measured at the LHC in the range $2.76\leq\sqrt{s}\leq 13$ TeV \cite{TOTEM0}. While the total, $\sigma_{tot}$, integrated elastic, $\sigma_{el}$ and inelastic, $\sigma_{in}$ cross sections, in general follow the expectations and extrapolations from lower energies, several new, unexpected features, challenging the standard Regge-pole model were discovered in elastic scattering. These are: 
	\begin{enumerate}
		\item low-$|t|$ structure  in $d\sigma/dt$ (the "break") \cite{totem83,TOTEM_rho};
		\item unexpectedly rapid rise of the forward slope $B(s,t=0)$ \cite{TOTEM0};
		\item surprisingly low value of the phase of the forward amplitude \cite{TOTEM_rho};
		\item absence of secondary dips/bumps in $d\sigma/dt$ \cite{TOTEM_rho,totem7}. 
	\end{enumerate}
	In this paper we analyze these and related phenomena within a Regge pole model, emphasizing their deviation from earlier trends and expectations, on the one hand and their discovery potential on the other hand. 
	
	The deviation of the diffraction cone from the exponential, colloquially called the "break" was first discovered in 1972 at the ISR \cite{Bar}. It immediately attracted attention of theorists, who interpreted it as manifestation of the pionic atmosphere surrounding the nucleon due to the two-pion loop in the $t$ channel \cite{LNC, AG, C-I1, C-I2}. The effect was not seen for 40 years 
	but it reappeared at the LHC, with statistics exceeding that of the ISR experiment, enabling now its more detailed study, in particular its origin and universality.
	
	The slope is defined as 
	\begin{equation}\label{Eq:slope}
	B(s,t\rightarrow 0)=\frac{d}{dt}\Biggl(\ln\frac{d\sigma}{dt}\Biggr)\bigg|_{t=0},
	\end{equation}
    where $\frac{d\sigma}{dt}$ is the elastic differential cross section.
	In the case of a single and simple Regge pole, the slope increases logarithmically with $s$:
	\begin{equation}\label{Eq:tslope}
	B(s,t)=B_0(t)+2\alpha'(t)\ln(s/s_0).
	\end{equation}
	
	Recall that at energies below the Tevatron, including those of SPS and definitely so at the ISR, secondary trajectories contribute significantly. At the LHC and beyond, we are at the fortunate situation where these non-leading contribution may be neglected. Apart from the pomeron, the odderon, critical at the dip \cite{JLL} may be present. 
	Remind also that a single Regge pole produces a monotonic $\ln s$ rise of the slope, discarded by the recent LHC data \cite{TOTEM0}. 
	
	The identification of the odderon is an important open problem for theory and phenomenology.
	In Ref. \cite{Ster} a nearly model-independent method to extract the odderon from $pp$ and
	$\bar pp$ data was suggested. The role of the odderon at the LHC is discussed in detail in the present paper. 
	
	A model-independent Levy expansion to describe the structure of the diffraction cone was suggested recently \cite{Levy} providing a statistically acceptable description of the differential cross-section of $pp$ and $\bar pp$ elastic
	scattering from ISR to LHC energies, but only with a fairly large number of Levy expansion coefficients.
	
	We rely on the dipole pomeron (DP) model to handle the observed phenomena. Being simple, it nevertheless reproduces the complicated diffractive structure of the differential cross section (dip-bump), enabling to trace the role of the odderon at the LHC.    
	
	The appearance of a the diffraction structure (dips and bumps) may be indicative also
	of the internal structure of the nuclei. The details of this structure may reveal the proton's structure: number of constituents (quarks), their size etc. Such a program was  put forward some time ago in Refs. \cite{Waka} and \cite{GV} and, recently by M. Csan\'ad {\it et al.} \cite{NCsCs}, who conclude that the observed dip-bump structure indicates the existence of diquarks inside the nucleons.

	In Sec.~\ref{Sec:DP} we present the dipole pomeron (DP) model (for a review see {\it e.g.} \cite{PEPAN}), the main tool in our analysis, with fits to high-energy $pp$ and $\bar pp$ observables (elastic differential cross section, total cross section and parameter $\rho$). 
	The DP pomeron model is the unique alternative to a simple pole, since higher order poles are not allowed by unitarity. The DP produces (logarithmically) rising cross sections even at unit intercept. A particularly attractive feature of the DP is the built-in mechanism of the diffraction pattern: a single minimum appears, followed by a maximum in the differential cross section, confirmed by the experimental data in a wide span of energies. In Sec.~\ref{sec:B} we 
	scrutinize the recently reported measurement of the forward slope, apparently exceeding the standard logarithmic rise typical of Regge-pole models. In Sec.~\ref{sec:rho} we argue that the widely disputed value $\rho(13~TeV)=0.1$ by itself cannot be considered of an odderon signal. Much more critical in this (odderon signal) respect is the dynamics of the dip and bump, discussed in Sec. \ref{sec:dip}. Our analyses is completed in Sec.~\ref{sec:relcontr}, where we quantify our statement about the negligible role of secondary trajectories at the LHC and the progressively increasing contribution from the odderon.
	
	\section{The dipole pomeron (DP) model}\label{Sec:DP}
	In our opinion, the Regge-pole model is the most adequate, although not unique way to analyze high-energy elastic hadron scattering.  Regge-pole models are attractive for being economic, especially in describing $C-$even and $C-$odd (the odderon!) contributions with a single set of free parameters. 
	
	The construction of any scattering theory consists of two stages: one first chooses an input amplitude ("Born term"), subjected to a subsequent unitarization procedure. Neither the input, nor the unitarization procedure are unique. In any case, the better the input, \textit{i.e.} closer to the true amplitude, the better are the  chances of the unitarization. The standard procedure is that of Regge-eikonal, {\it i.e.} when the eikonal is identified with a simple Regge-pole input.
	
	A possible alternative to the simple Regge-pole model as input is a double pole (double pomeron pole, or simply dipole pomeron, DP) in the angular momentum $(j)$ plane. It has a number of advantages over the simple pomeron Regge pole. In particular, it produces logarithmically rising cross sections already at the "Born" level. 
	
	In this section we prepare the ground by introducing the model amplitude and fitting its parameters to the data. At the LHC, the pomeron dominates, however, to be consistent with the lower energy data, particularly those from the ISR, we include also two secondary reggeons. The odderon, pomeron's odd-$C$ counterpart is also included in the fitting procedure.

	As already mentioned, the pomeron is a dipole in the $j-$plane
	\begin{eqnarray}\label{Pomeron}
	& &A_P(s,t)={d\over{d\alpha_P}}\Bigl[{\rm e}^{-i\pi\alpha_P/2}G(\alpha_P)\Bigl(s/s_{0P}\Bigr)^{\alpha_P}\Bigr]= \\ \nonumber
	& &{\rm e}^{-i\pi\alpha_P(t)/2}\Bigl(s/s_{0P}\Bigr)^{\alpha_P(t)}\Bigl[G'(\alpha_P)+\Bigl(L_P-i\pi
	/2\Bigr)G(\alpha_P)\Bigr].
	\end{eqnarray}
	Since the first term in squared brackets determines the shape of the cone, one fixes
	\begin{equation} \label{residue} G'(\alpha_P)=-a_P{\rm
		e}^{b_P[\alpha_P-1]},\end{equation} where $G(\alpha_P)$ is recovered
	by integration. Consequently the pomeron amplitude Eq.~(\ref{Pomeron}) may be rewritten in the following "geometrical" form (for details see \cite{PEPAN} and references therein):
	\begin{equation}\label{GP}
	A_P(s,t)=i{a_P\ s\over{b_P\ s_{0P}}}[r_{1P}^2(s){\rm e}^{r^{2}_{1P}(s)[\alpha_P-1]}-\varepsilon_P r_{2P}^2(s){\rm e}^{r^2_{2P}(s)[\alpha_P-1]}],
	\end{equation} 
	where $r_{1P}^2(s)=b_P+L_P-i\pi/2$, $r_{2P}^2(s)=L_P-i\pi/2$, \mbox{$L_P\equiv\ln(s/s_{0P})$}. The pomeron trajectory, in its simplest version is linear:
	\begin{equation}\label{Ptray}
	\alpha_P\equiv \alpha_P(t) = 1+\delta_P+\alpha'_{P}t.
	\end{equation}
	
	A remarkable property of the DP pomeron, noticed by R. Phillips \cite{Phillips}, see also the Appendix in Ref.~\cite{Fort} is that it scales, {\it i.e.} reproduces itself against unitarity corrections. Really, the DP amplitude Eq.~(\ref{GP}) in the impact parameter representation is Gaussian. Eikonalization, in $n-$th approximation means, roughly speaking raising the impact parameter amplitude to the power $n$, leaving its functional form intact (scaling) and, since the parameters anyway are fitted to the data, the fitted model is close to unitary. 
	
	In earlier versions of the DP, to avoid conflict with the Froissart bound, the intercept of the pomeron was fixed at $\alpha(0)=1$. However later it was realized that the logarithmic rise of the total cross sections provided by the DP may not be sufficient to meet the data, therefore a supercritical intercept was allowed for. From the earlier fits to the data the value $\delta=\alpha(0)-1\approx0.04,$ half of Landshoff's value \cite{Land} follows. This is understandable: the DP promotes half of the rising dynamics, thus moderating the departure from unitarity at the "Born" level (smaller unitarity corrections).
	
	We assume that the odderon contribution is of the same form as that of the pomeron, implying the relation $A_O=-iA_P$ and different values of adjustable parameters (labeled by subscript ``$O$''): 
	\begin{equation}\label{Odd}
	A_O(s,t)={a_O\ s\over{b_O\ s_{0O}}}[r_{1O}^2(s){\rm e}^{r^2_{1O}(s)[\alpha_O-1]}-\varepsilon_O r_{2O}^2(s){\rm e}^{r^2_{2O}(s)[\alpha_O-1]}],
	\end{equation}
	where $r_{1O}^2(s)=b_O+L_O-i\pi/2$, $r_{2O}^2(s)=L_O-i\pi/2$, \mbox{$L_O\equiv\ln(s/s_{0O})$} and the trajectory
	\begin{equation}\label{Eq:Otray}
	\alpha_O\equiv \alpha_O(t) = 1+\delta_O+\alpha'_{O}t.
	\end{equation}
	
	Secondary reggeons are parametrized in a standard way \cite{KKL, KKL1}, with linear Regge trajectories and exponential residua. The $f$ and $\omega$ reggeons are the principal non-leading contributions to $pp$ or $\bar p p$ scattering:
	\begin{equation}\label{Reggeon1}
	A_f\left(s,t\right)=a_f{\rm e}^{-i\pi\alpha_f\left(t\right)/2}{\rm e}
	^{b_ft}\Bigl(s/s_0\Bigr)^{\alpha_f\left(t\right)},
	\end{equation}
	\begin{equation}\label{Reggeon2}
	A_\omega\left(s,t\right)=ia_\omega{\rm e}^{-i\pi\alpha_\omega\left(t\right)/2}{\rm e}
	^{b_\omega t}\Bigl(s/s_0\Bigr)^{\alpha_\omega\left(t\right)},
	\end{equation}
	with $\alpha_f\left(t\right)=0.703+0.84t$ and
	$\alpha_{\omega}\left(t\right)=0.435+0.93t$. 
	
	While the Pomeron and $f$-reggeon have positive C-parity, thus they enter to the scattering amplitude with the same
	sign in $pp$ and $\bar pp$ scattering, the Odderon and $\omega$-reggeon have negative C-parity,
	entering in $pp$ and $\bar pp$ scattering with opposite signs. The complete scattering amplitude used in our fits is:
	\begin{equation}\label{Eq:Amplitude}
	A\left(s,t\right)_{pp}^{\bar pp}=A_P\left(s,t\right)+A_f\left(s,t\right)\pm\left[A_{\omega}\left(s,t\right)+A_O\left(s,t\right)\right].
	\end{equation}
	
	We use the norm where
	\begin{equation}\label{norm}
	\sigma_{tot}(s)={4\pi\over s}Im A(s,t=0)\Bigl.\ \ {\rm and}\ \
	{d\sigma_{el}\over{dt}}(s,t)={\pi\over s^2}|A(s,t)|^2 \  .
	\end{equation}
	
	The parameter $\rho(s)$, the ratio of the real and imaginary part of the forward scattering amplitude is 
	\begin{equation}\label{eq:rho}
	\rho(s)=\frac{Re A(s,t=0)}{Im A(s,t=0)}.
	\end{equation}
	
	The free parameters of the model defined by the formulas Eqs.~(\ref{GP}-\ref{eq:rho}) were fitted simultaneously to the following dataset:
	
	\begin{itemize}
		\item TOTEM 7 TeV elastic $pp$ differential cross section data \cite{totem7} in the interval $0.35\leqslant|t|\leqslant2.5$ GeV$^2$;
		\item SPS 546 and 630 GeV elastic $p\bar p$ differential cross section data \cite{Battiston:1983gp,Bernard:1986ye,Bozzo:1985th} in the interval $0.5\leqslant|t|\leqslant2.2$ GeV$^2$;
		\item $pp$ and $p\bar p$ total cross section and $\rho$ parameter data \cite{TOTEM0,TOTEM_rho,totem7,totem72,PDG,totem81,totem82,Auger} in the interval $20\leqslant\sqrt{s}\leqslant57000$ GeV.
	\end{itemize}

    In the present paper, by using the DP model we focus on the LHC energies, scrutinizing the role of the odderon. To properly extract the odderon we must include the $pp$ and $p\bar p$ differential cross section data in the dip-bump and the "shoulder" (in $\bar pp$) regions. When this paper was submitted, only 7 TeV $pp$ differential cross section data were published in the "dip-bump" region. To account for the “shoulder” effect, we included also the SPS data.
    
    The published data on LHC TOTEM 7 TeV proton-proton differential cross section \cite{totem7} are compiled from two subsequent measurements below and above $-t=0.37$ GeV$^2$. The inclusion of the low-$|t|$ data would deteriorate the fit statistics. It was discussed also in Ref. \cite{NCsCs}. The situation is the same in case of the SPS 546 GeV data, where the low-$|t|$ and high-$|t|$ measurements match near $-t=0.5$ GeV$^2$ \cite{Bozzo:1985th}.
    
    In order enable proper investigation of the weights of the different amplitude components (see Sec.~\ref{sec:relcontr}) at LHC energies, we include in our analysis non-leading secondary reggeons whose parameters are known from previous fits \cite{JLL}. Here we let free only their normalization parameters ($a_f$ and $a_\omega$), that due to the use in the fits of the low-energy total cross section and the $\rho$-parameter data enable to match the leading high-energy  and non-leading low-energy contributions in the model.
    
	The fit was done using MINUIT2 and MIGRAD algorithms. The values of fitted parameters and the fit statistics are shown in Table~\ref{tab:parameters}.
	To optimize the fit, following the ideas of Ref.\cite{RuizArriola:2017kqs} we have removed 63 outlying ({\it i.e.} lying outside the trend) data points on the differential cross section.  This procedure reduces $\chi^2/NDF$ from 2.4 to 1.4 and the $NDF$ value from 222 to 159. We have fixed also the pomeron normalization parameter ($a_P$), otherwise MINUIT2 fails in finding the minimum, producing an invalid fit.

	\begin{table}[H]	
		\centering
		\caption{Values of the parameters fitted to $pp$ and $p\bar p$ data on elastic differential cross section, total cross section and the ratio~$\rho$.}
		\begin{tabular}{|ccc|ccc|}
			\hline \hline
			\multicolumn{1}{|c}{}&\multicolumn{2}{c}{Pomeron}&\multicolumn{2}{c}{Odderon}&\multicolumn{1}{c|}{} \\
			\hline 
			&$a_P~[\sqrt{\rm mbGeV^2}]$ & $360$ (fixed) &$a_O~[\sqrt{\rm mbGeV^2}]$&$1.772\pm0.113$& \\
			&$b_P$ & $4.143\pm0.152$  &$b_O$ &$0.8981\pm0.0062$&         \\
			&$\delta_P$ & $0.02908\pm0.00056$ &$\delta_O$&$0.2746\pm0.0049$&  \\
			&$\alpha'_{P}~[{\rm GeV}^{-2}]$ & $0.5079\pm0.0028$  &$\alpha'_{O}~[{\rm GeV}^{-2}]$ &$0.2310\pm0.0018$&  \\
			&$\varepsilon_P$ & $0.2823\pm0.0154$  &$\varepsilon_O$&$1.311\pm0.004$&  \\
			&$s_{0P}~[{\rm GeV}^2]$ & $100$ (fixed) &$s_{0O}~[{\rm GeV}^2]$ & $100$ (fixed)&\\ \hline \hline
			\multicolumn{1}{|c}{}&\multicolumn{2}{c}{Reggeons}&\multicolumn{2}{c}{Fit statistics}&\multicolumn{1}{c|}{} \\ \hline
			&$a_f~[\sqrt{\rm mbGeV^2}]$ & $-20.09\pm0.17$&&&\\
			&$b_f~[{\rm GeV}^{-2}]$  & $4$ (fixed) &$\chi^2$&223.1&\\
			& $a_{\omega}~[\sqrt{\rm mbGeV^2}]$ & $10.64\pm0.64$&$NDF$&159&\\
			&$b_{\omega}~[{\rm GeV}^{-2}]$ & $15$ (fixed) &$\chi^2/NDF$&1.4&\\
			&$s_0~[{\rm GeV}^2]$  & $1$ (fixed)&&&\\ \hline \hline
		\end{tabular}
		\label{tab:parameters}
	\end{table}
	
	\section{Elastic, inelastic and total cross sections}\label{Sec:El}
	
\begin{figure}[H]
		\centering
		\includegraphics[scale=0.25]{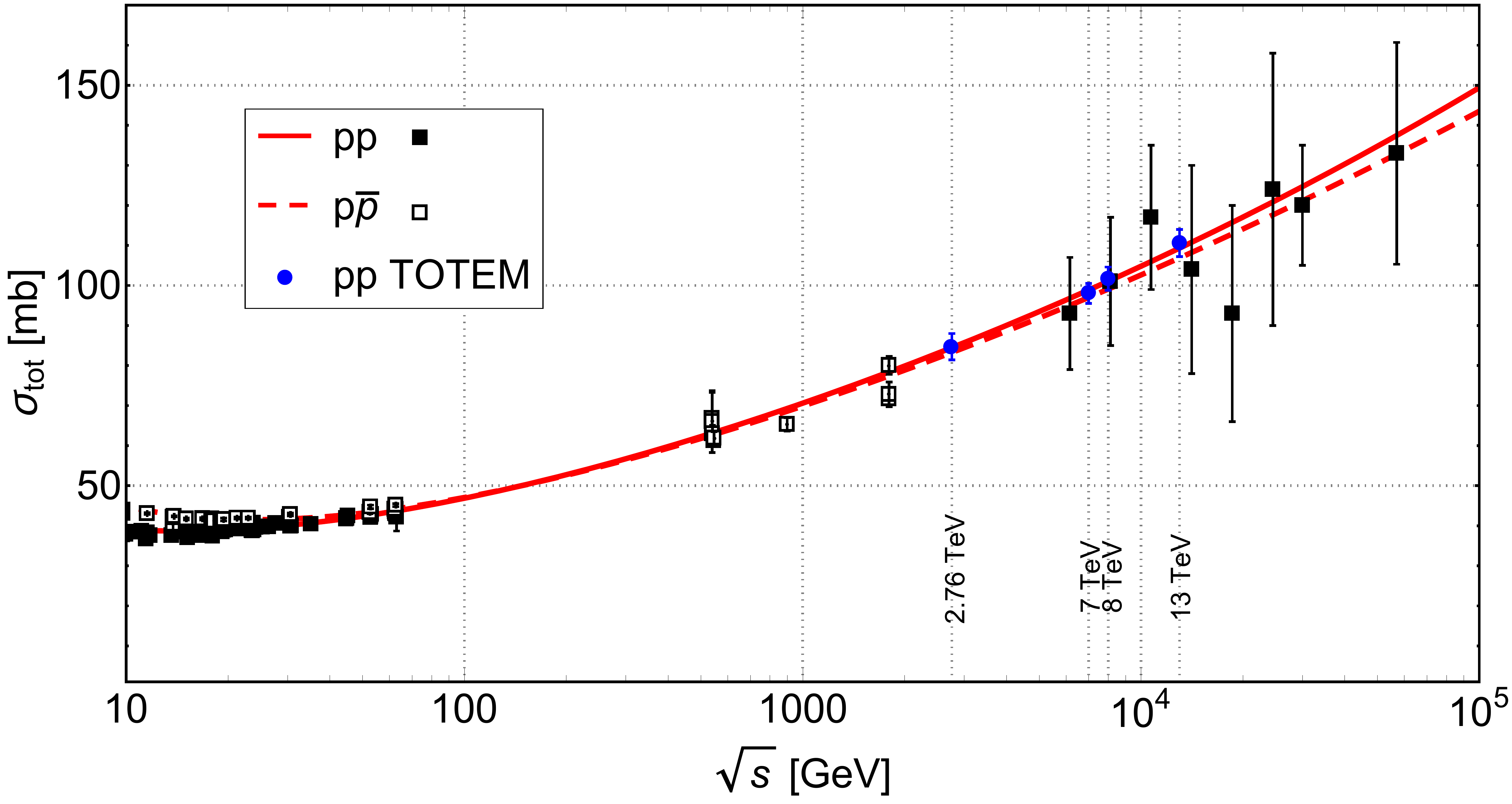}
		\vspace{-0.5cm}
		\caption{Fits to $pp$ and $p\bar p$ total cross section data \cite{TOTEM0,totem72,PDG,totem81,Auger} using the model Eqs.~(\ref{GP}-\ref{eq:rho}).}
		\label{Fig:sigma}
	\end{figure}

	Fig.~\ref{Fig:sigma} shows the results of our fits to $pp$ and $p\bar p$ total cross section data \cite{TOTEM0,totem72,PDG,totem81,Auger}.

\begin{figure}[H] 
		\centering
		\includegraphics[scale=0.25]{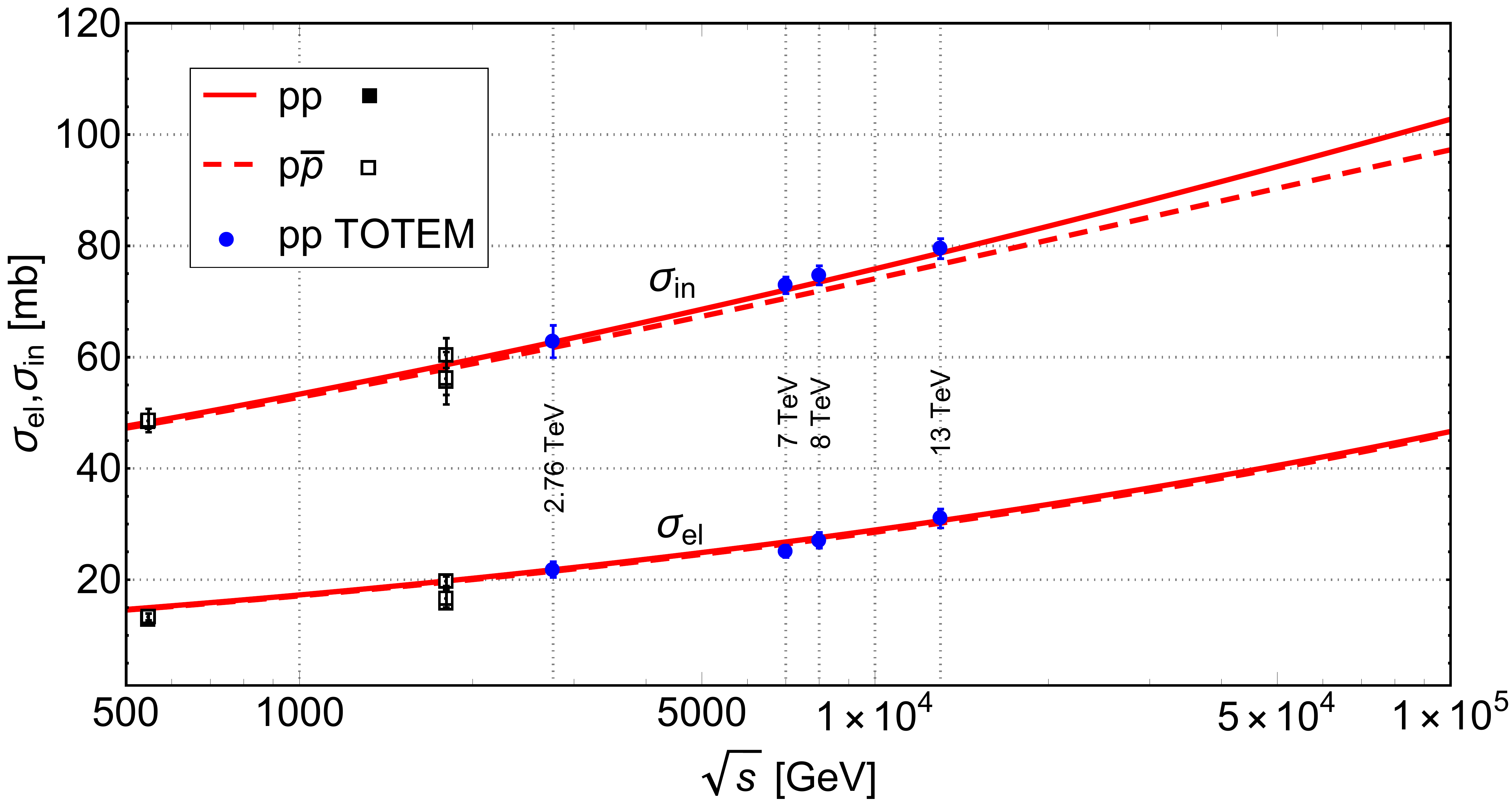}
		\vspace{-0.5cm}
		\caption{Calculated elastic and inelastic $pp$ and $p\bar p$ cross sections using Eqs.~(\ref{eq:el}-\ref{eq:inel}) compared to the data \cite{TOTEM0,totem72,PDG,totem81}.}
		\label{Fig:sigma_in_el}
	\end{figure}

	The elastic cross section $\sigma_{el}(s)$ is calculated by integration
	\begin{equation}\label{eq:el}
	\sigma_{el}(s)=\int_{t_{min}}^{t_{max}}\frac{d\sigma}{dt}(s,t)\, dt,
	\end{equation}
	whereupon
	\begin{equation}\label{eq:inel}
	\sigma_{in}(s)=\sigma_{tot}(s)-\sigma_{el}(s). 
	\end{equation}
	Formally, $t_{min}=-s/2$ and $t_{max}=t_{threshold}$, however since the integral is saturated basically by the first cone, we set $t_{max}=0$ and $t_{min}=-1$ GeV$^2$. The results are shown in Fig.~\ref{Fig:sigma_in_el}.
		
	One can see from Fig.~\ref{Fig:sigma} and Fig.~\ref{Fig:sigma_in_el} that the model reproduces reasonably the trends observed by the TOTEM Collaboration, with $\delta_P=0.02902>0$. 
	
	\section{The phase $\rm \rho(s, t)$}\label{sec:rho}
	
Extraction of the phase from Coulomb-nuclear interference is an important problem by itself, going beyond the present study. Here we use the $t=0$ value published by TOTEM. The merit of our model is that it reproduced the $t$ dependence of the phase beyond $t=0$, not accessible from the experiment.	
	
	The recent measurement of the phase $\rho(13~TeV)=0.09\pm0.01$ (or $\rho(13~TeV)=0.1\pm0.01$) \cite{TOTEM_rho} is vividly discussed in the literature. The above data point lies well below the expectations (extrapolations) from lower energies, although this should not be dramatized. The flexibility of the odderon parametrization leaves room for perfect fits to this data point simultaneously with the total cross section (see below). More critical is the inclusion of non-forward data, both for $pp$ and $\bar pp$ especially around the dip region, to which the odderon is sensitive! 
	
	This result has important consequences both for TOTEM and calculations of the matter distribution in the proton (so-called hollowness), discussed in paper [Broniowski] that appeared after the present one was submitted for publication. 
	
\begin{figure}[H]
		\centering
		\includegraphics[scale=0.25]{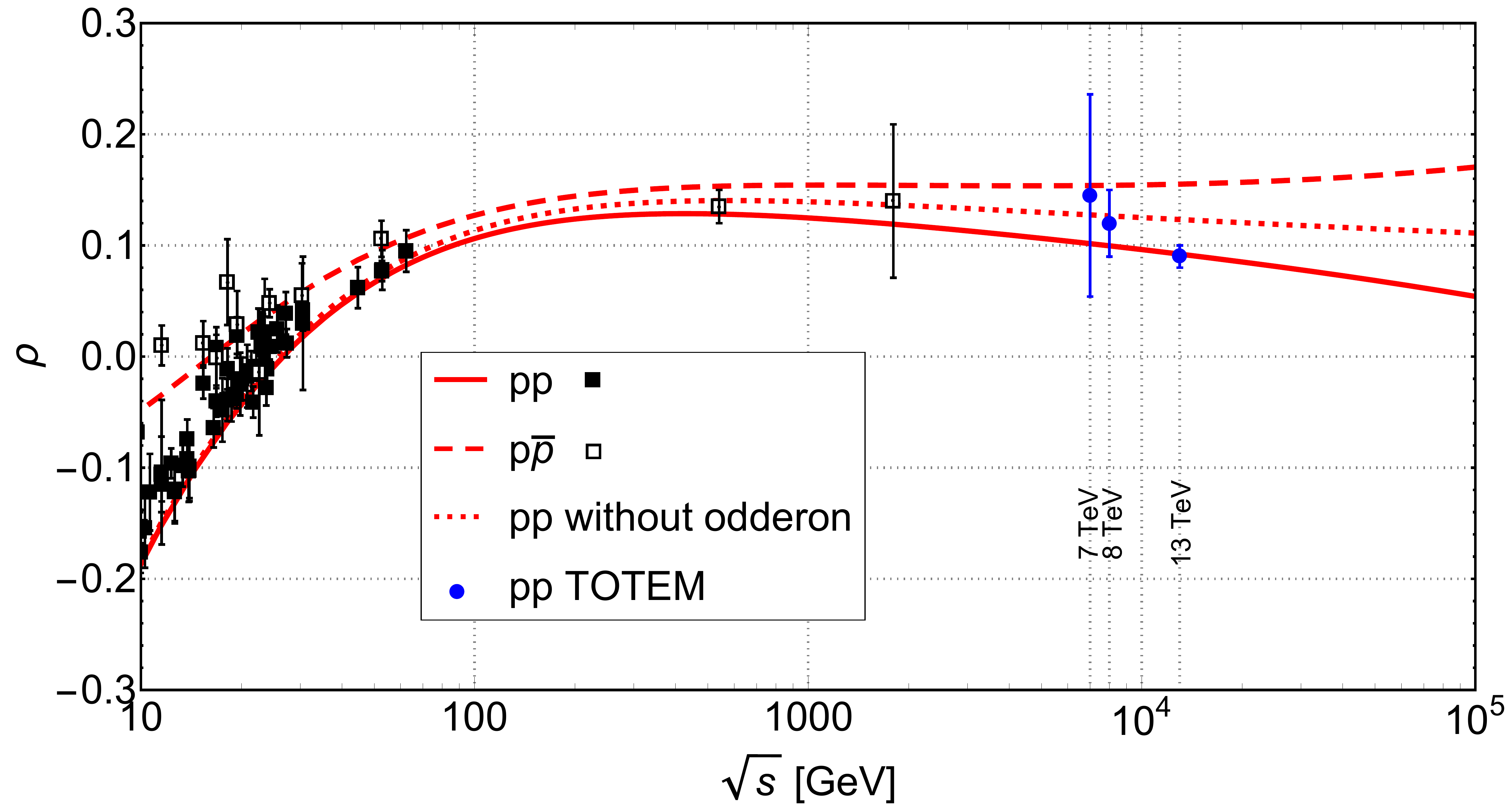}
		\caption{Fits to $pp$ and $p\bar p$ ratio $\rho$ data \cite{TOTEM_rho,totem7,PDG,totem82} using the model Eqs.~(\ref{GP}-\ref{Eq:Amplitude}) and Eq.~(\ref{eq:rho}).}
		\label{Fig:rho}
	\end{figure}

	Fig.~\ref{Fig:rho} shows the results of our fits to $pp$ and $p\bar p$ $\rho$-parameter data \cite{TOTEM_rho,totem7,PDG,totem82}. Our model fits simultaneously the new TOTEM measurement on the total cross section and the parameter $\rho$ \cite{TOTEM_rho}.
	
\begin{figure}[H]
		\centering
		\includegraphics[scale=0.25]{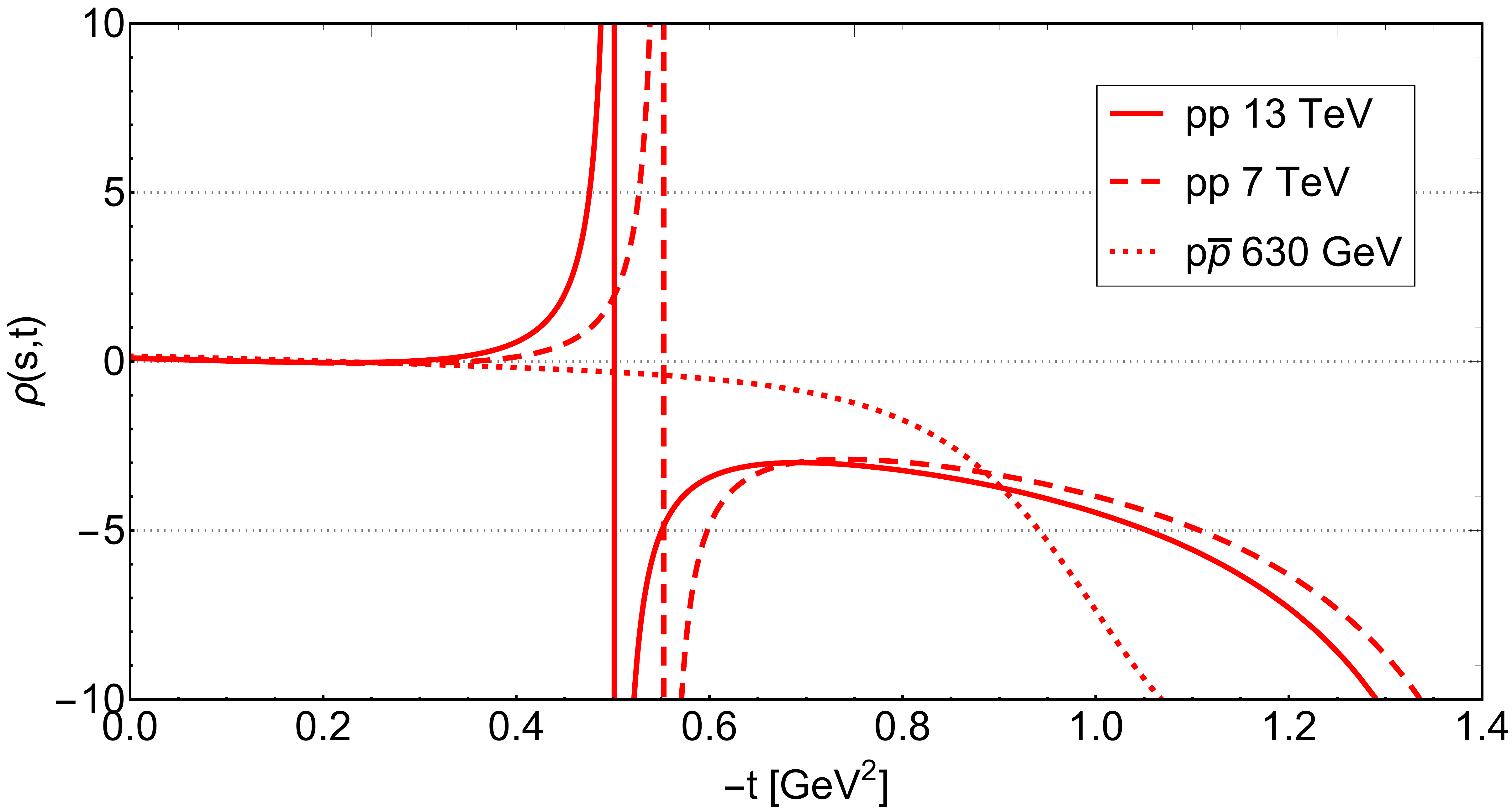}
		\vspace{-0.5cm}
		\caption{The t-dependent $\rho$ parameter at several energies calculated with the model Eqs.~(\ref{GP}-\ref{Eq:Amplitude}) and Eq.~(\ref{eq:rho}) (when $t\neq0$).}
		\label{Fig:rhot}  
	\end{figure}

\begin{figure}[H]
		\centering
		\includegraphics[scale=0.25]{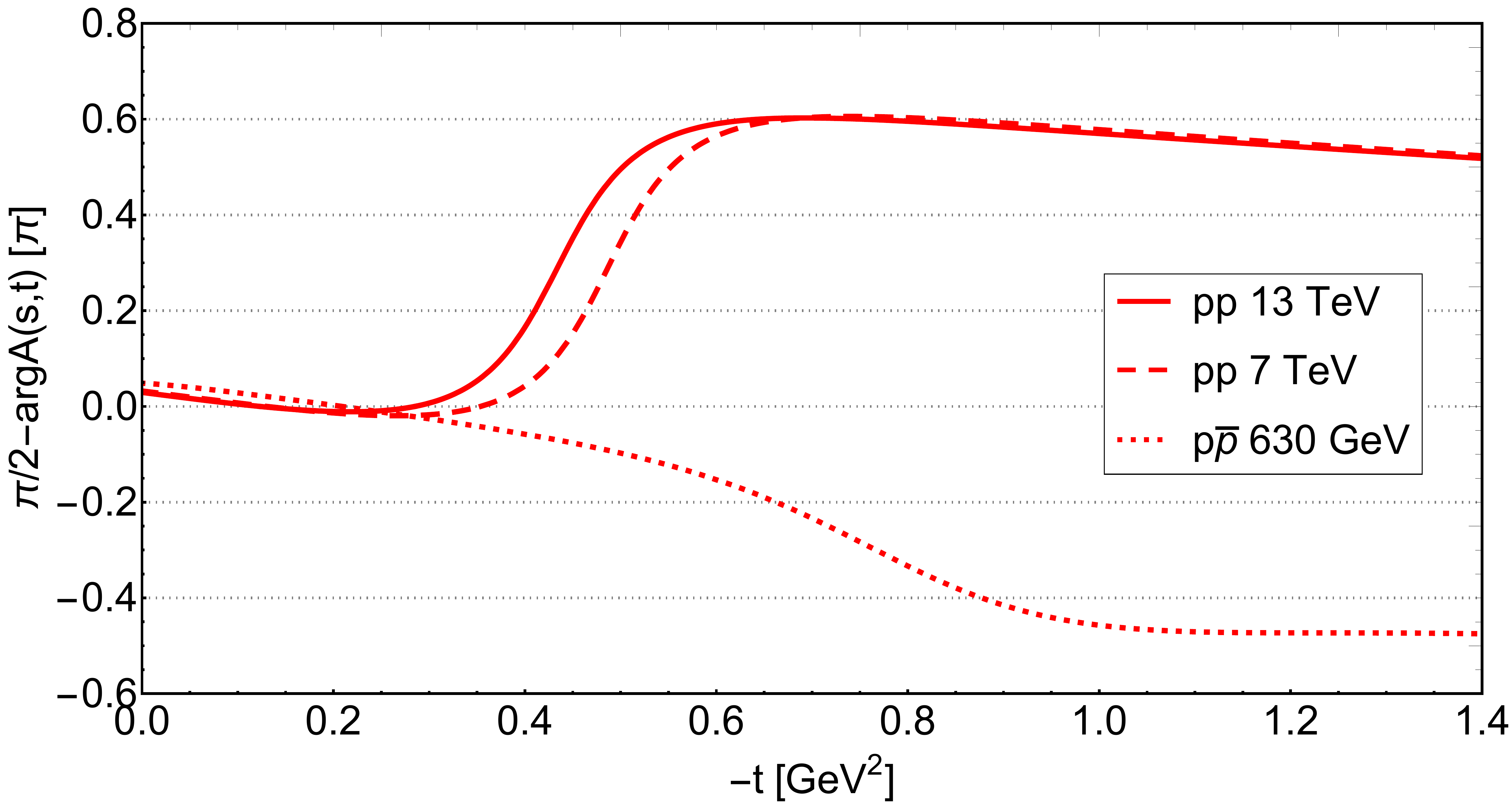}
		\caption{The t-dependent hadronic phase at several energies calculated with the model Eqs.~(\ref{GP}-\ref{Eq:Amplitude}).}
		\label{Fig:phaset}
	\end{figure}

	
	As seen in Fig.~\ref{Fig:rho}, the case without the odderon (shown as a dotted line) does not provide a description for the new 13 TeV data point. However, we found that neglect of the oddereon has no significant effect on the description of the new TOTEM total cross section measurements.   
	We calculated also the t-dependence of the $\rho$-parameter and the hadronic phase defined as $\phi(s,t)=\pi/2-\arg A(s,t)$ at several energies for $pp$ and $\bar pp$ scattering shown in Fig.~\ref{Fig:rhot} and Fig.~\ref{Fig:phaset}. The hadronic phase may be important in studies of the impact parameter amplitude \cite{BJASz}.
	
	\section{The slope $\rm B(s,t=0)$}\label{ssec:B}\label{sec:B}
	
	The slope of the cone $B(s,t)$ is not measured directly, instead it is extracted from the data on the directly measurable differential cross sections within certain bins in $t$. Therefore, the primary sources are the cross sections or the scattering amplitude fitted to these cross sections.
	
	For example, at the SPS Collider the slope was measured at $\sqrt{s}=540$ GeV \cite{Burq} in two $t$-bins with the result
	$$B=13.3\pm 1.5\ {\rm GeV}^{-2}\ \ \  (0.15<|t|<0.26)\ {\rm GeV}^2, $$
	$$B=17.2\pm 1.0\ {\rm GeV}^{-2}\ \ \  (0.05<|t|<0.18)\ {\rm GeV}^2. $$ 
	
	While in Regge-pole models the rise of the  total cross sections is regulated by the hardness of the Regge pole (here, the pomeron), the slope $B(s)$ in case of a single and simple Regge pole is always logarithmic. Deviation (acceleration) may arise from more complicated Regge singularities, the odderon and/or from unitarity corrections.
	
	With the model introduced in Sec.~\ref{Sec:DP} and its fitted parameters in hand, we calculated the $pp$ and $p\bar p$ elastic slope $B(s)$ using Eq.~(\ref{Eq:slope}). The result is shown in Fig.~\ref{Fig:B}. Details of the calculations 
	may be found in the Appendix. 
	
	By using the parametrization
	\begin{equation}\label{eq:slope2}
	B(s)=b_0+b_1\ln(s/s_0)+b_2\ln^2(s/s_0)
	\end{equation}
	we performed a fit to the $pp$ and $p\bar p$ elastic slope data in the energy region $546\leqslant\sqrt{s}\leqslant13000$ GeV. The best fit was achieved when the three lowest among the six measured slopes at 546 GeV were excluded. The result, with $b_0=13.75$, $b_1=-0.358$, $b_2=0.0379$ and $\chi^2/NDF=1.97$ is shown in Fig.~\ref{Fig:B}.
	
\begin{figure}[H]	
		\centering
		\includegraphics[scale=0.25]{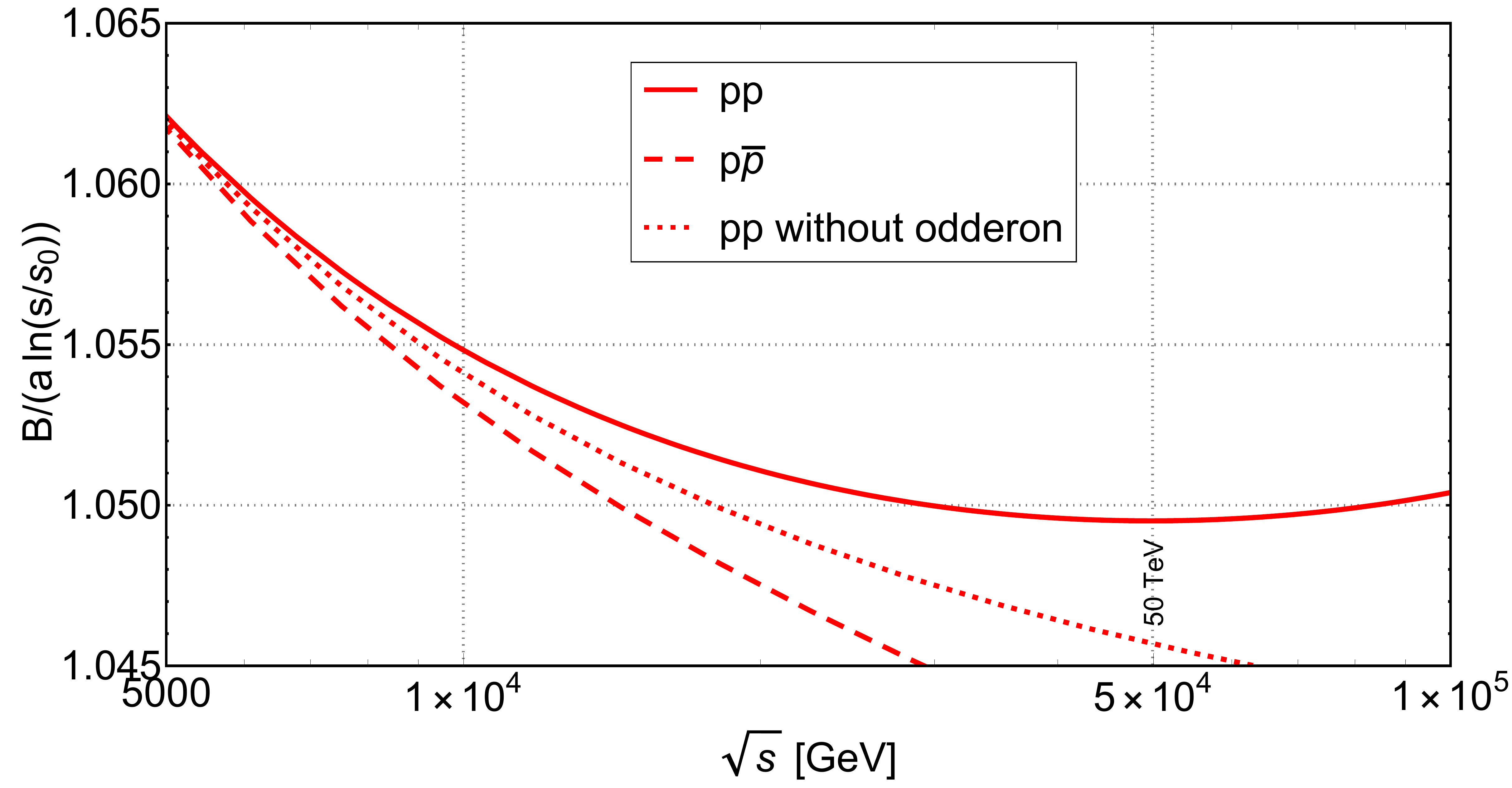}
		\caption{The ratio $B(s)/(a\ln(s/s_0))$ calculated from the fitted model Eqs.~(\ref{GP}-\ref{norm}) using Eq.~(\ref{Eq:slope}).}
		\label{Fig:Bo}
	\end{figure}

	To see better the effect of the odderon and the deviation of $B(s)$ from its "canonical" logarithmic form, we show in Fig.~\ref{Fig:Bo} its  "normalized" shape, $B(s)/(a\ln(s/s_0))$ setting $a=1$ GeV$^{-2}$ and $s_0=1$ GeV$^{2}$. A similar approach was useful in studies \cite{totem83,Break3,Break1,Break2} of the fine structure (in $t$) of the diffraction cone.
	
	In Fig.\ref{Fig:Bo} the "normalized" $pp$ curve starts rising from $\sqrt{s}\approx50$ TeV, indicating that the slope increases faster than $\ln s$. The dipole pomeron alone, without the odderon (dotted curve in Fig.~\ref{Fig:Bo}) at the "Born" level, fitting to the data on elastic, inelastic and total cross section, does not reproduce the irregular behavior of the forward slope observed at the LHC. Remarkably, the inclusion of the odderon promotes a faster than $\ln s$ rise of the elastic slope $B(s)$ beyond the LHC energy region.
	
	
	\begin{figure}[H] 
		\centering
		\includegraphics[scale=0.25]{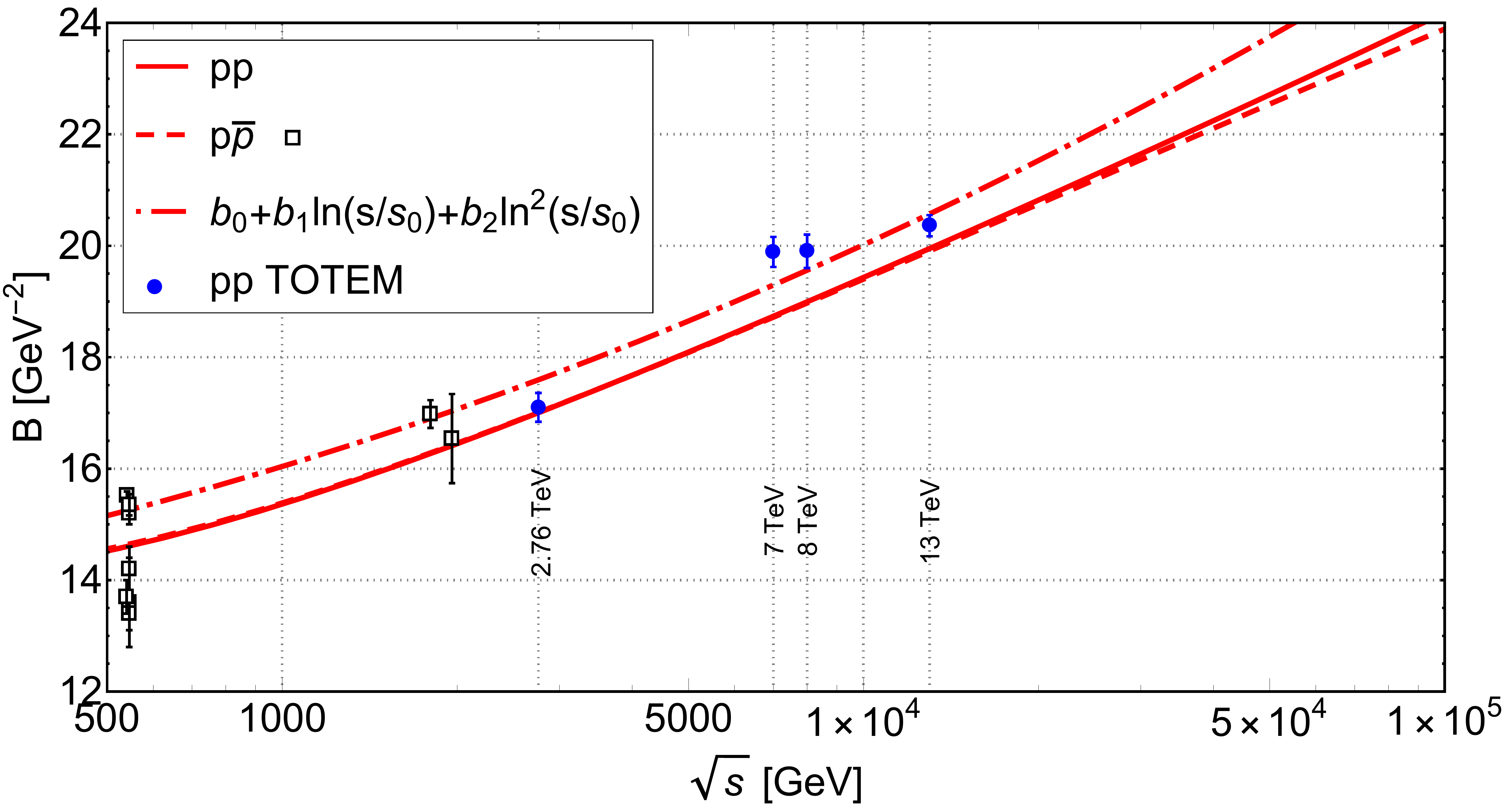}
		\caption{$pp$ and $p\bar p$ elastic slope $B(s)$ a) calculated  from the fitted model Eqs.~(\ref{GP}-\ref{norm}) using Eq.~(\ref{Eq:slope}) and b) fitted with the parametrization Eq.~(\ref{eq:slope2}).}
		\label{Fig:B}
	\end{figure}
	
Neither of the mentioned fits is satisfactory, indicating an open problem for the theory. A possible solution may be found within the BSW model \cite{BSW1, BSW2, BSW3}, in which the non-trivial behavior of the slope was predicted \cite{BSW6} well before the TOTEM measurements.	
		
	\section{The diffraction minimum and maximum (dip-bump)}\label{sec:dip}
	The most sensitive (crucial) test for any model of elastic scattering is the well-known dip-bump structure in the differential cross section. It was measured in a wide range of energies and squared momenta transfers. None of the existing models was able to predict the position and dynamics of the dip for (especially when both $pp$ and $\bar pp$ data are included). The first LHC measurements (at $7$~TeV) \cite{totem7} clearly demonstrated their failure.
	
	Recently the TOTEM Collaboration made public \cite{Nemes} new, preliminary data on elastic differential cross section at highest LHC energy $13$ TeV extending up to $t=-3.8$ GeV$^2$. The main message from these data is that the second cone is smooth, structureless. This finding calls for the revision of models in which the dip is created by unitarization resulting in interference between single and multiple scattering or, alternatively by eikonal corrections, generating multiple diffraction minima and maxima. 
	
	Recall that the dip is deepening in $pp$ scattering at the ISR energies (up to a certain energy, whereupon the dynamics seems to change). Instead, in $\bar pp$ scattering a shoulder appears at the place of the expected dip, which may be an indirect evidence in favour of the odderon, filling in the dip. A direct test of the odderon implies simultaneous (at the same energy) measurement of both $pp$ and $\bar pp$ cross sections. Their difference would indicate the presence of an odd-$C$ exchange. This happened only once \cite{Breakstone}, before the shut-down of the ISR, where the difference showed unambiguously the presence of an odd $C$ contribution, that, however can be attributed both to the $\omega$ and/or odderon exchange.  
	
\begin{figure}[H]
	\centering
	\includegraphics[scale=0.245]{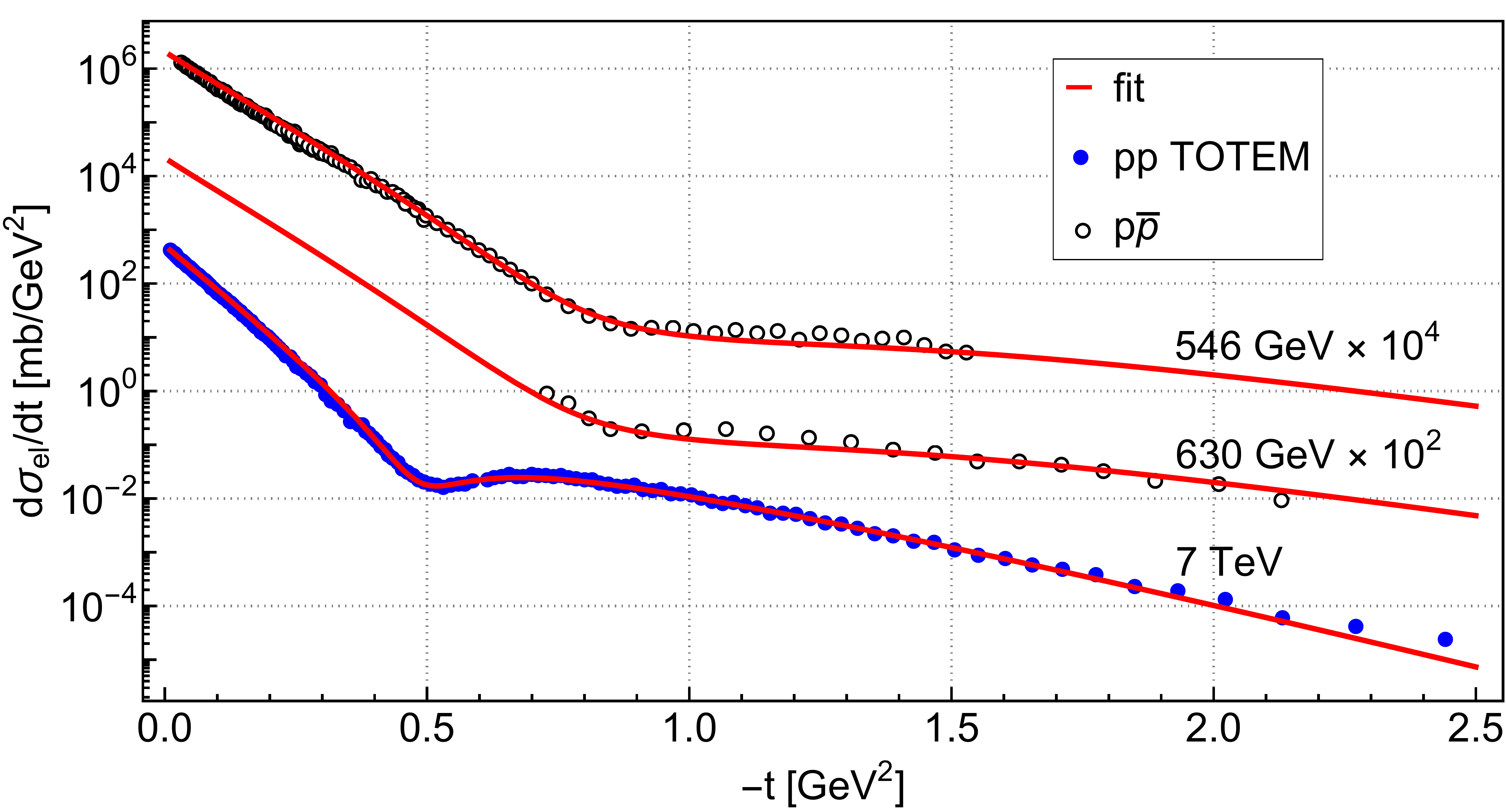}%
	\caption{Results of the fit for $pp$ and $p\bar p$ differential cross section data \cite{totem7,Battiston:1983gp,Bernard:1986ye,Bozzo:1985th} using the model Eqs.~(\ref{GP}-\ref{norm}).}
	\label{Fig:dsigma}
\end{figure}

	Before going into details, let us remind that the present DP model predicts 
	$(d\sigma_{el}/dt)_{min}\sim 1/L$, $(d\sigma_{el}/dt)_{max}\sim L$ and consequently  
	$\frac{(d\sigma_{el}/dt)_{max}}{(d\sigma_{el}/dt)_{min}}\sim L^2,$ where $L=\ln s$, in the case of a single pomeron contribution. The addition of the odderon, due to its opposite $C$ parity, destroys the dip in case of $\bar pp$ scattering (degrading to a "shoulder" at the dip position). In $pp$ the odderon contributes as given by Eq.~(\ref{Eq:Amplitude}), however the overall effect depends on the details or the parametrization. We have performed a fit of the differential cross section in the dip region including both the pomeron and odderon. The result for $pp$ and $p\bar p$ differential cross sections, using Eqs.~(\ref{GP})-(\ref{norm}) is shown in Fig.~\ref{Fig:dsigma}.

\begin{figure}[H]
		\centering
		\includegraphics[scale=0.25]{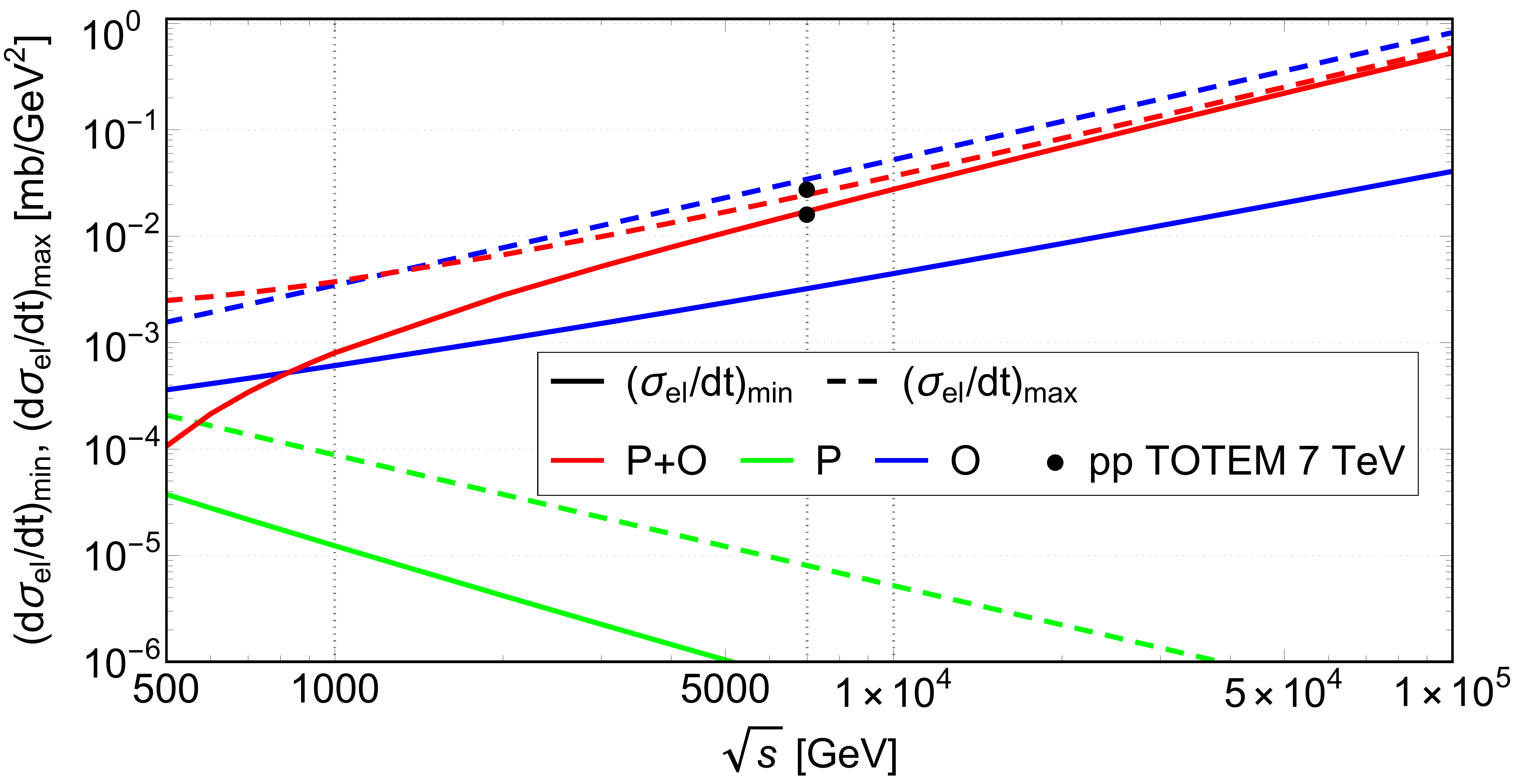}%
		\caption{Energy dependence of the maximum and minimum of the $pp$ diffraction cone calculated from the fitted model.}
		\label{Fig:maxmin}
	\end{figure}

\begin{figure}[H]
		\centering
		\includegraphics[scale=0.25]{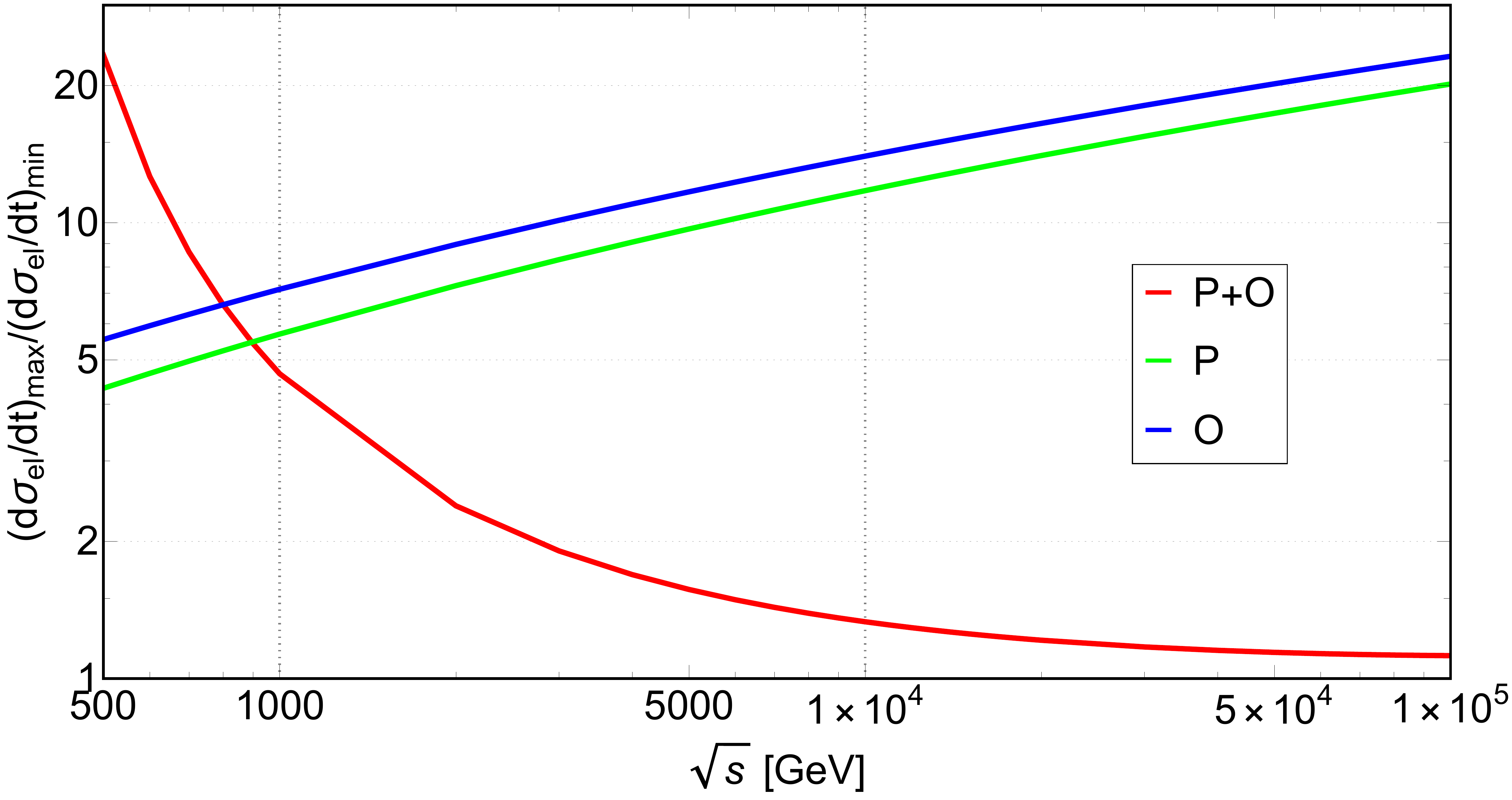}%
		\caption{Energy dependence of the ratio of the maximum and minimum of the $pp$ diffraction cone calculated from the fitted model.}
		\label{Fig:maxpermin}
	\end{figure}

	Next we study the energy dependence of the minimum and maximum of $pp$ diffraction cone \textit{i.e.} the behavior of $(d\sigma_{el}/dt)_{min}(s)$, $(d\sigma_{el}/dt)_{max}(s)$ and their ratio $\frac{(d\sigma_{el}/dt)_{max}(s)}{(d\sigma_{el}/dt)_{min}(s)}$. The mentioned quantities were calculated numerically from the fitted model and they are plotted in Fig.~\ref{Fig:maxmin} and Fig.~\ref{Fig:maxpermin}. As it will be shown in Sec.~\ref{sec:relcontr}, in the dip-bump region at high energies ($\sqrt{s}\gtrsim500$ GeV) the secondary reggeons can be neglected completely, so we have a chance to see the contributions of the pomeron and the odderon alone in the evolution of the minimum and maximum.

	We recall that in the DP model with unit pomeron intercept, the maximum rises as $L$ while the minimum deepens also as $L$, resulting in a $L^2$ increase of their ratio. However, this simple picture is obscured by: a) the presence of the odderon, whose role increases with $t$ and b) the larger than unity intercepts of both the pomeron and odderon.
	
	One can see from Fig.~\ref{Fig:maxpermin} that in the ratio $\frac{(d\sigma_{el}/dt)_{max}(s)}{(d
		\sigma_{el}/dt)_{min}(s)}$, the pomeron and the odderon components separately increase monotonically producing a monotonically deepening minimum, although the ratio $\frac{(d\sigma_{el}/dt)_{max}(s)}{(d\sigma_{el}/dt)_{min}(s)}$ decreases. 
	The result depends however on the choice of the odderon, for which many options exist. The problem was studied also by O. Selyugin using the HEGS model \cite{Selyugin} predicting that the $\frac{(d\sigma_{el}/dt)_{max}(s)}{(d\sigma_{el}/dt)_{min}(s)}$ ratio increases at LHC energies. 
As stressed in the previous paragraph, the behaviour of the ratio $\frac{(d\sigma_{el}/dt)_{max}(s)}{(d\sigma_{el}/dt)_{min}(s)}$, explicit for the separate pomeron and odderon contributions, becomes complicated by their interference and depends on the fitted parameters.   	
	
	\section{The "break"}\label{sec:beak}
The non-exponential behavior of the low-$|t|$ $pp$ differential cross section was confirmed by recent measurements by the TOTEM Collaboration at the CERN LHC, first at $8$ TeV (with a significance greater than 7$\sigma$) \cite{totem83} and subsequently at $13$ TeV \cite{TOTEM_rho}. 

At the ISR the "break" (in fact a smooth curve) was illustrated by plotting the local slope \cite{Bar} 
\begin{equation}
B(t)=\frac{d}{dt} \ln (d\sigma/dt ) \,
\end{equation}
for several $t$-bins at fixed values of $s$. 
Unlike the ISR case, the TOTEM quantifies the deviation from the exponential by normalizing the measured cross section to a linear exponential form \cite{totem83,TOTEM_rho}. 

The normalized form, used by TOTEM is: 
\begin{equation} \label{Eq:norm}
R(t)=\frac{d\sigma/dt}{d\sigma/dt_{ref}}-1,
\end{equation}
where $d\sigma/dt_{ref}=Ae^{Bt}$, with $A$ and $B$ constants determined from a fit to the experimental data. 

The observed "break" can be identified \cite{LNC, C-I1, C-I2,Break3,Break1,Break2}  with two-pion exchange (loop) in the $t$-channel. As shown by Barut and Zwanziger \cite{Barut}, $t$-channel unitarity constrains the Regge trajectories near the threshold, $t\rightarrow t_0$ by
\begin{equation} \label{Eq:Barut}
Im\, \alpha(t)\sim (t-t_0)^{\Re e\, \alpha(t_0)+1/2},
\end{equation} 
where $t_0$ is the lightest threshold, $4m_{\pi}^2$ in the case of the vacuum quantum numbers (pomeron or $f$ meson). 

\begin{figure}[H] 
	\centering
	\includegraphics[scale=0.25]{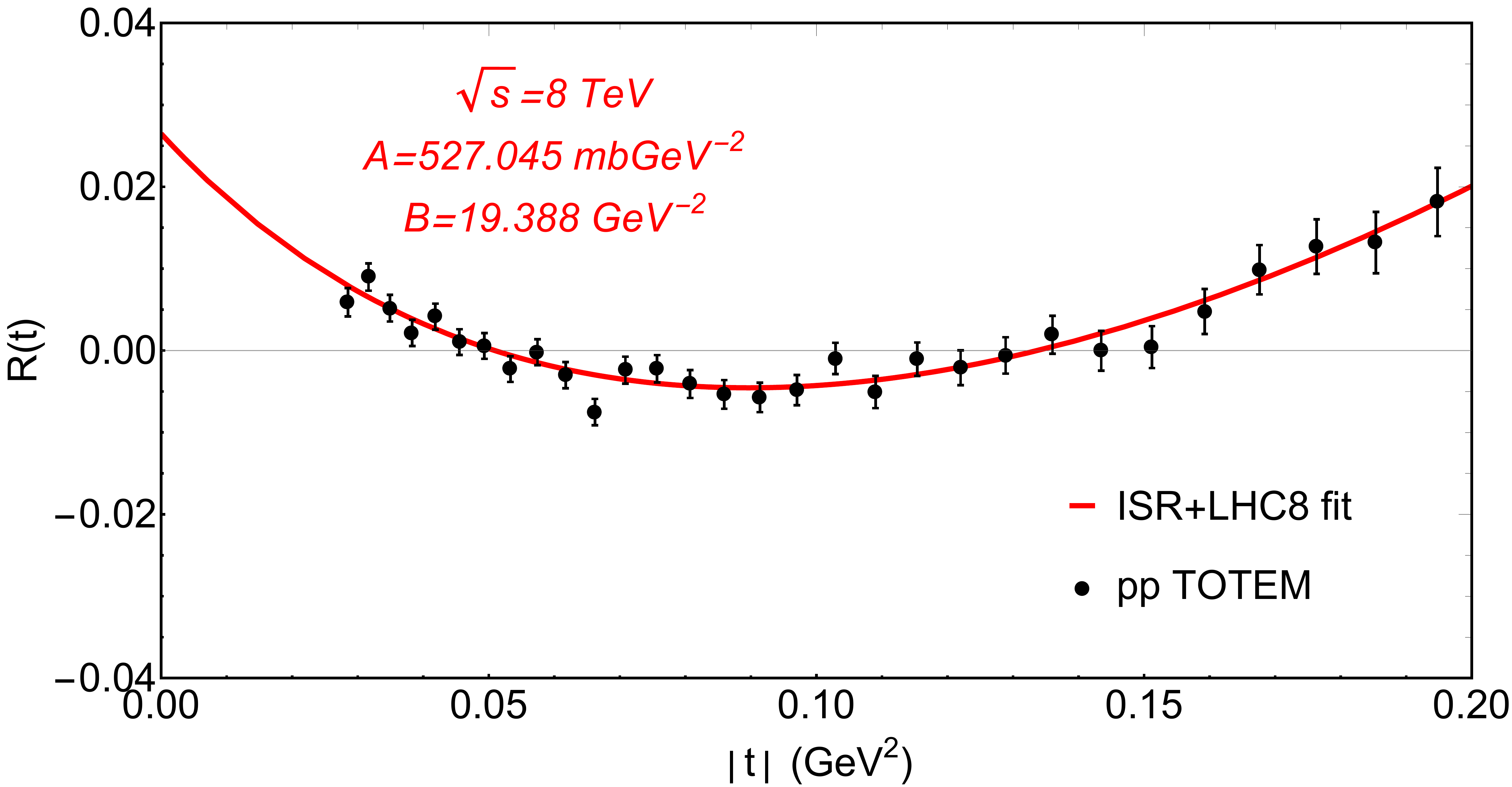}
	\vspace{-0.5cm}
	\caption{Normalized differential cross section $R(t)$ calculated from low-$|t|$ 8 TeV TOTEM data \cite{totem83} using Eq.~(\ref{Eq:norm}). This figure is from Ref.~\cite{Break3}.}
	\label{Fig:Rratio1}
\end{figure}

\begin{figure}[H] 
	\centering
	\includegraphics[scale=0.25]{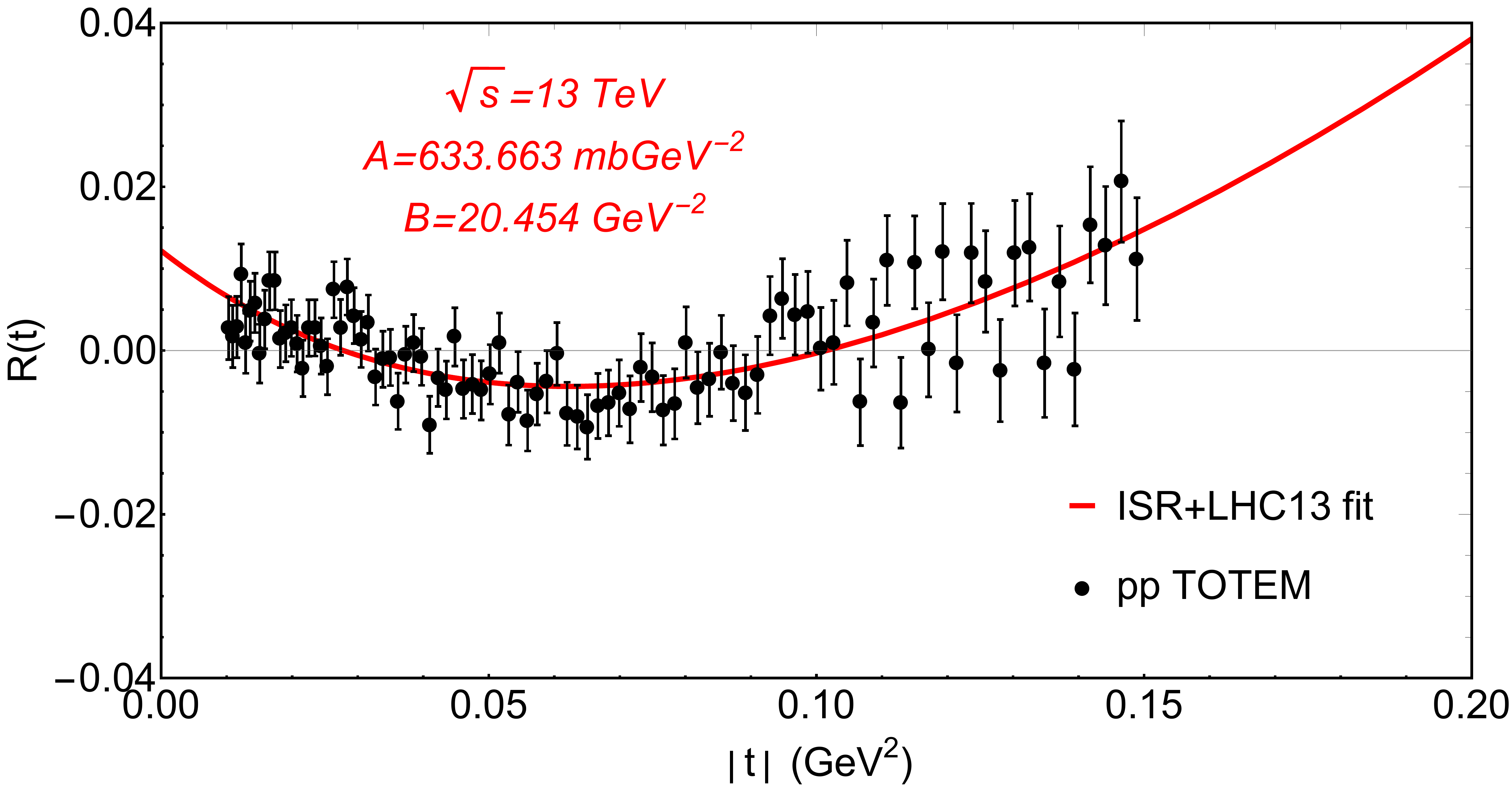}
	\vspace{-0.5cm}
	\caption{Normalized differential cross section $R(t)$ calculated from low-$|t|$ 13 TeV TOTEM data \cite{TOTEM_rho} using Eq.~(\ref{Eq:norm}). This figure is from Ref.~\cite{Break3}.}
	\label{Fig:Rratio2}
\end{figure}

In the above fits, focusing on the reole of the odderon in the "dip-bump" dynamimcs, we did not include the low-$|t|$ region with the "break". For the sake of  completeness, below we show the main results of the analysis of the non-exponential diffraction cone based on our recent paper Ref.~\cite{Break3}. Fig.~\ref{Fig:Rratio1} and Fig.~\ref{Fig:Rratio2} show our description (in normalized form) to the "break" measured at 8 and 13 TeV. The contribution from the two-pion exchange is mimicked by the $4m_{\pi}^2$ threshold modifying the linear trajectory Eq. (\ref{Ptray}) as: 
	\begin{equation}
	\alpha_P\equiv \alpha_P(t) = 1+\delta_P+\alpha'_{P}t-\alpha_1\sqrt{4m_{\pi}^2-t}.
	\end{equation}
More details, including the values of the parameters may be found in Ref. \cite{Break3}. 

These results re-confirm the earlier finding that the ``break" can be attributed the presence of two-pion branch cuts in the Regge parametrization. 

In Ref.~\cite{Break3}. among others, two aspects of the phenomenon were investigated, namely: 1) to what extent is the "break" observed recently at the LHC is a "recurrence"  of that seen at the ISR (universality)? 2) what is the relative weight of the Regge residue (vertex) compared to the trajectory (propagator) in producing the "break"? We showed that the deviation from a linear exponential of the $pp$ diffraction cone as seen at the ISR and at the LHC are of similar nature: they appear nearly at the same value of $t\approx -0.1$~GeV$^2$, similar shape and size, and may be fitted by similar $t$-dependent function. Furthermore, we found that the Regge residue and the pomeron trajectory have nearly the same weight and importance.

Note that while the dip moves towards lower $|t|$ with energy, this is not necessarily true for the "break", whose origin, nature and energy dependence is quite different from that of the dip \cite{Break2}.
	
	\section{Relative contribution from different components of the amplitude}\label{sec:relcontr}
	
\begin{figure}[H]	
		\centering
		\includegraphics[scale=0.25]{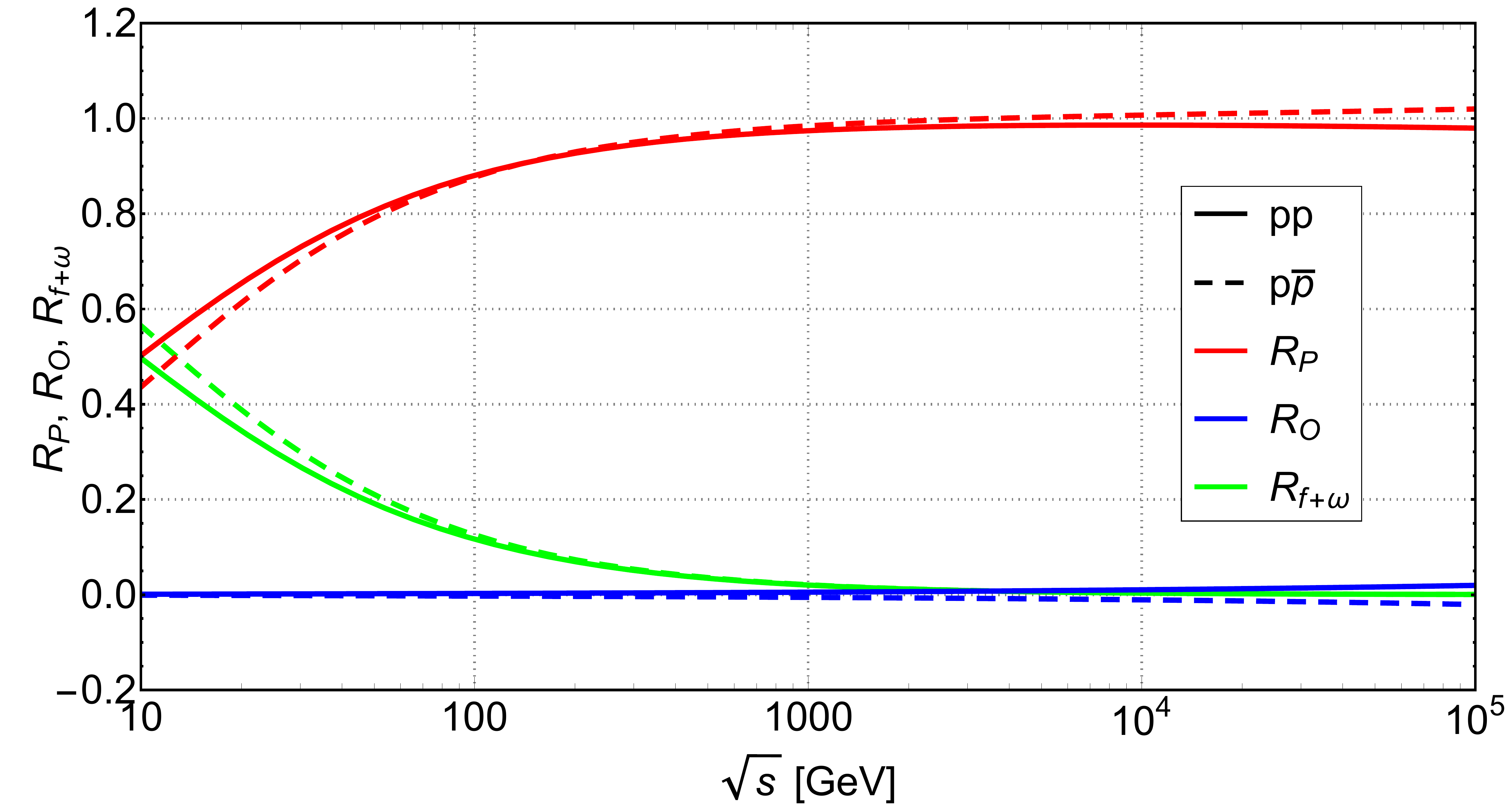}
		\caption{Relative contribution from different components of the amplitude to $pp$ and $p\bar p$ total cross-sections  calculated from the model, Eqs.~(\ref{GP}-\ref{Eq:Amplitude}), Eq.~(\ref{eq:relative}).}
		\label{Fig:r1}
	\end{figure}

	Within the framework of the model Eqs.~(\ref{GP}-\ref{Eq:Amplitude}), we calculated the relative contribution from the different components of the amplitude
	\begin{equation}\label{eq:relative}
	R_i(s)={{Im A_i(s,t=0)\over{Im A(s,t=0)}}},
	\end{equation}
	to the $pp$ and $p\bar p$ total cross-sections, where $i=f+\omega$ for the relative weight of the reggeons, $i=P$ for the relative weight of the pomeron and $i=O$ for the relative weight of the odderon. The result is shown in Fig.~\ref{Fig:r1}.
	
	One can see from Fig.~\ref{Fig:r1} that at "low" energies (typically 10 GeV) the contribution from reggeons and the pomeron are nearly equal, but as the energy increases the pomeron takes over and at the same time the importance of the odderon is slightly growing.
	
	Such a discrimination (between the different contributions of the components of the amplitude) is more problematic in the non-forward direction, where the real and imaginary parts of various components of the scattering amplitude behave in a different way and the phase can not be controlled experimentally.
	
\begin{figure}[H]	
		\centering
		\includegraphics[scale=0.245]{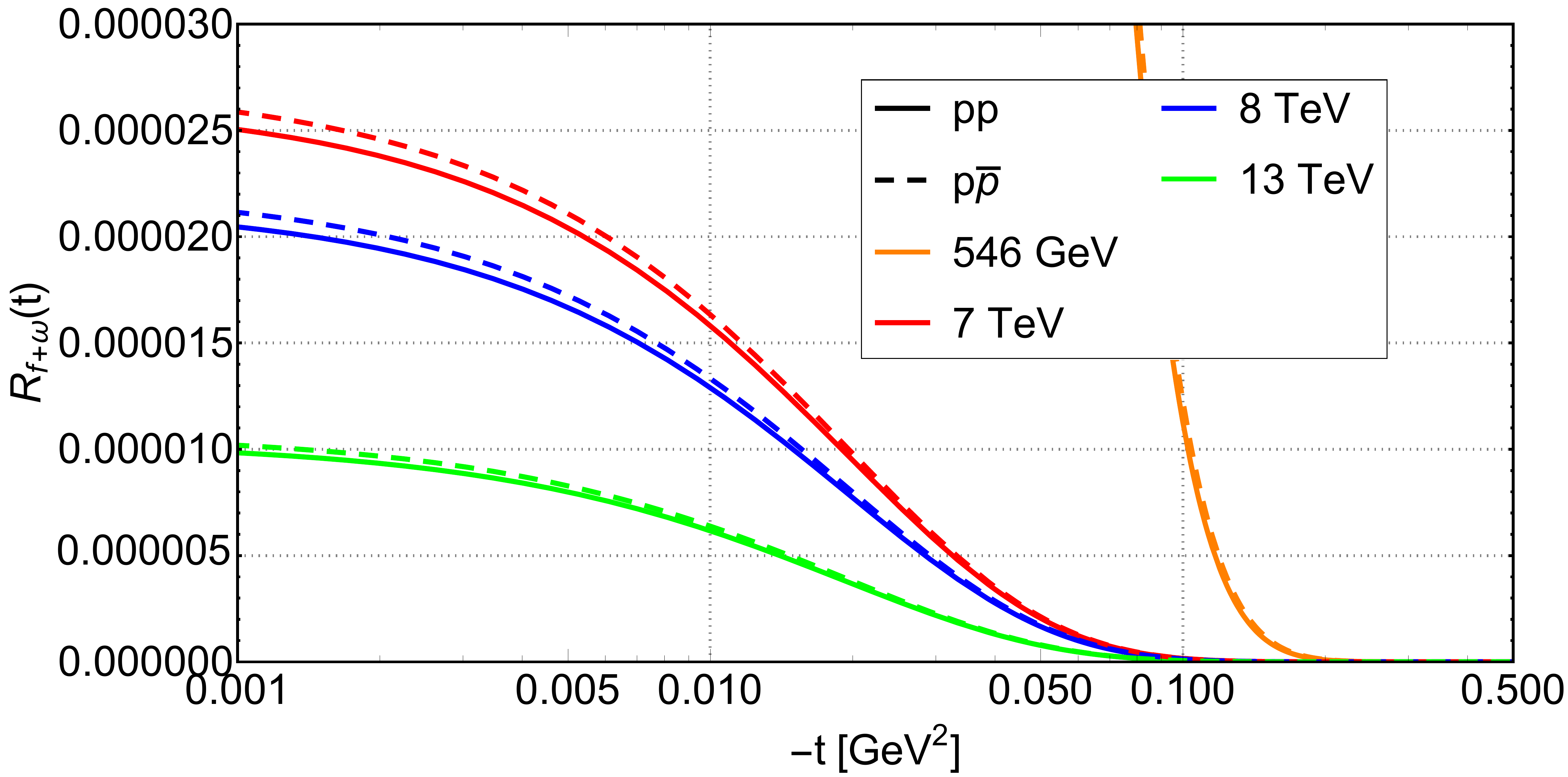}
		\caption{Relative contribution from secondary reggeons to the $pp$ and $p\bar p$ differential cross-sections calculated from the model, Eqs.~(\ref{GP}-\ref{Eq:Amplitude}), Eq.~(\ref{eq:relativet}).}
		\label{Fig:r2}
	\end{figure}

	We calculate the relative contributions of different components of the amplitude for non-forward scattering ${(t\neq 0)}$:
	\begin{equation}\label{eq:relativet}
	R_i(s,t)={{\left|A_i(s,t)\right|^2\over{\left|A(s,t)\right|^2}}}.
	\end{equation}
	The relative contribution from secondary reggeons $R_{f+\omega}(s,t)$ versus $-t$ at $546$ GeV, $7, 8$ and $13$ TeV is shown in Fig.~\ref{Fig:r2}.
	One can see that the role of the secondary reggeons rapidly decreases with increasing $|t|$ values. 
	
\begin{figure}[H]	
		\centering
		\includegraphics[scale=0.235]{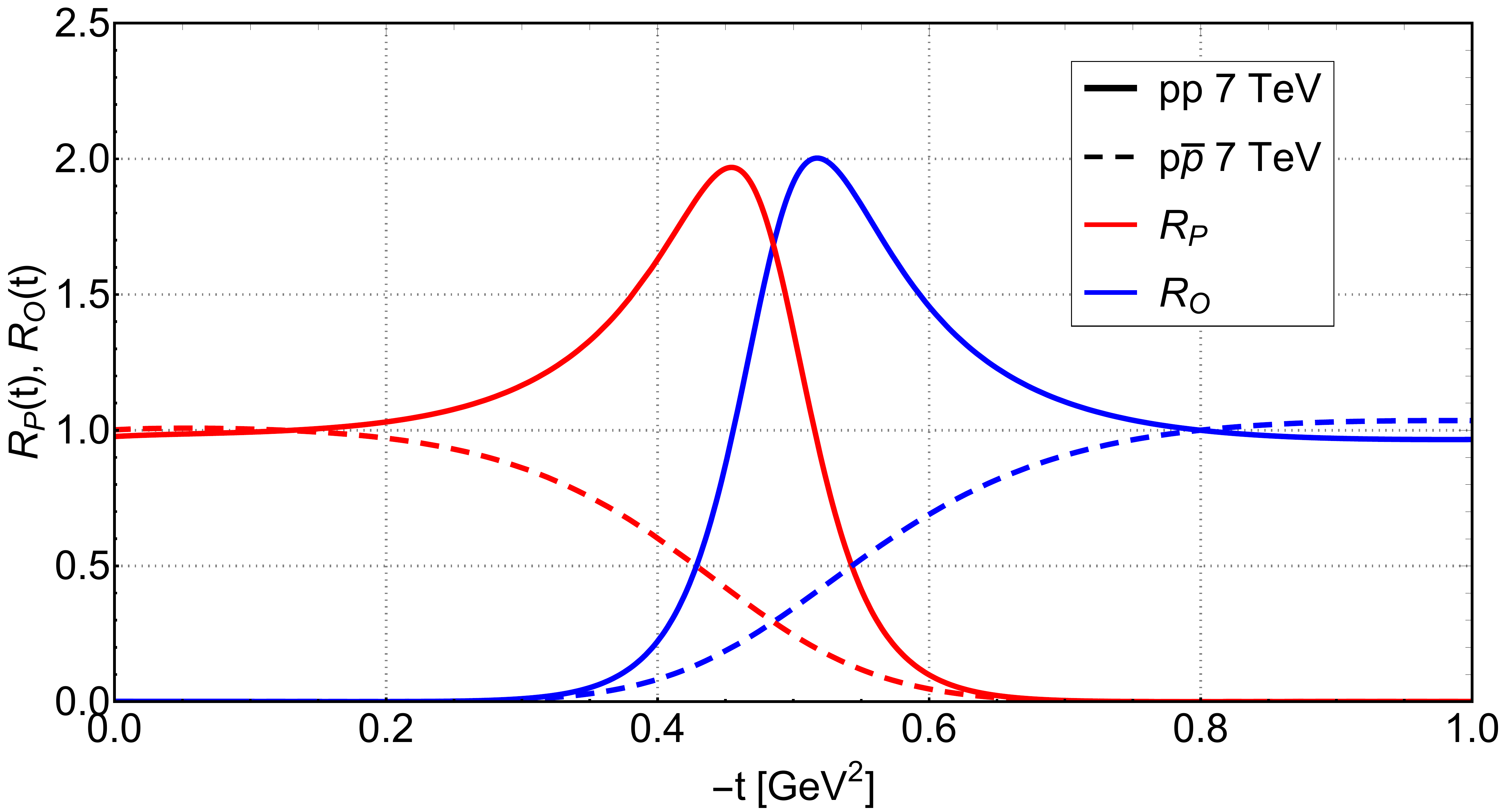}
		\caption{Relative contribution from the pomeron and from the odderon to the $pp$ and $p\bar p$ differential cross-sections at 7 TeV calculated from the model, Eqs.~(\ref{GP}-\ref{Eq:Amplitude}), Eq.~(\ref{eq:relativet}).}
		\label{Fig:r3}
	\end{figure}

	Furthermore, we calculated the relative importance in $t$ of the pomeron $R_{P}$ and of the odderon $R_{O}$ at 7 TeV. The result is shown in Fig.~\ref{Fig:r3}. One can see that at low $|t|$ values the pomeron completely dominates, 
	then around the dip-bump region the pomeron-odderon importance is about 50-50 \% and, finally at higher $|t|$ values the odderon takes over.
	
	\section*{Conclusions}
	We conclude that:
	
	1. The "break" is a universal feature of the forward cone. Regge-pole models interpolate this effect from the ISR energies up to those of the LHC. The "break" is due to the non-linear behavior of the pomeron trajectory and the non-exponential Regge residue, both resulting from a threshold singularity in the amplitude due to $t-$channel unitarity (two-pion loop in the $t$-channel).
	
	2. A single diffraction minimum (and maximum) in $d\sigma/dt$ is produced by a particular interference between a single and double pomeron poles. Unitarization, for example eikonalizaiton produces multiple dips and bumps. 
	
	
	3. The observed non-monotonic rise of the slope $B(s,t)$ at the LHC is incompatible with a single pomeron pole. Even the combination of a simple and double pole (DP) cannot accommodate for the accelerated rise of $B(s)$ at the LHC. The acceleration  of the forward slope $B(s)$, reported by TOTEM, however may be affected by the odderon. Thus, the odderon may play and important role also in the behavior of $B(s)$ at high energies, as shown in Fig. \ref{Fig:Bo}. 
	
	4. The low value $\rho(13~TeV)=0.1$ reported recently by TOTEM was not predicted however it was adjusted (together with the total cross section data) in several papers published after the appearance of the experimental data. 
	In our opinion, $\rho(13~TeV)=0.1$ alone cannot be considered as a "proof" of the odderon, whose existence can be hardly questioned. Critical may become simultaneous fits to $\sigma_{tot}(s),\ \rho(s)$ and differential cross sections including the dip-bump region at various energies both for $pp$ and $\bar pp$ scattering. 
	
	5. An interesting and important result of our analysis is that the role of the odderon is increasing with increasing squared momenta transfer $|t|$. The odderon starts dominating beyond the dip-bump region. This   observation is in accord with that of Donnachie and Landshoff (DL), see \cite{Land}, according to which the second slope (that beyond the dip-bump) is dominated by three-gluon exchange. Note that DL's "3-gluon" ansatz predicts energy independence of the diffraction cone at large $|t|$ while the TOTEM data at 13 TeV clearly shows the evident decrease as compare with ISR.  
	
	\section*{Acknowledgments}
	We thank Tam\'as Cs\"org\H{o} and Frigyes Nemes for useful discussions and correspondence. L.~J. was supported by the Ukrainian Academy of Sciences' program "Structure and dynamics of statistical and quantum systems". The work of N.~Bence and I.~Szanyi was supported by the "M\'arton \'Aron Szakkoll\'egium" program.  
	
	\section*{Appendix}
	The slope $B(s)$, calculated from Eq.~(\ref{Eq:slope}) with the norm Eq.~(\ref{norm}) and the amplitude Eq.~(\ref{Eq:Amplitude}) is:
	\begin{equation}
	B(s)=\frac{a(s)+b(s)L+c(s)L^2+d(s)L^3}{e(s)+f(s)L+g(s)L^2},
	\end{equation}
	where
	\begin{eqnarray} \label{Eq:abcdi}
	&a(s)&=\sum_{i=1}^{10}a_is^{k_i}, \ \ \ b(s)=\sum_{i=1}^{10}b_is^{k_i}, \\ \nonumber &c(s)&=\sum_{i=4}^{10}c_is^{k_i}, \ \ \ d(s)=\sum_{i=8}^{10}d_is^{k_i}, \\ \nonumber
	&e(s)&=\sum_{i=1}^{10}e_is^{k_i}, \ \ \ f(s)=\sum_{i=4}^{10}f_is^{k_i}, \\ \nonumber &g(s)&=\sum_{i=8}^{10}g_is^{k_i}.
	\end{eqnarray}
	The parameters $a_i$, $b_i$, $c_i$, $d_i$, $e_i$, $f_i$, $g_i$, and $k_i$ are related to the parameters in Eqs.~(\ref{Reggeon1}-\ref{Eq:Otray}).
	By neglecting the oddereon, the terms with $i=5$, $7$, $9$ and $10$ will be eliminated.  
    By keeping only secondary reggeons, terms with $i=4$, $6$ and $8$ disappear and one gets:
	\begin{equation}
	B(s)=\frac{a(s)+b(s)L}{e(s)},
	\end{equation}
	where $i=1$, $2$ and $3$.
	The expression for the slope Eq.~(\ref{Eq:slope}) with the single pomeron (or odderon) Eq.(\ref{GP})  (with trajectory Eq.~(\ref{Ptray}) reduces to:
	\begin{equation} \label{Eq:DPB}
	B(s)=\frac{2\alpha_{1P}(a+bL+cL^2+dL^3)}{e+fL+gL^2},
	\end{equation}
	where the parameters $a$, $b$, $c$, $d$, $e$, $f$ are energy-independent. They may be easily expressed in terms of those Eq.(\ref{GP}):
	\begin{eqnarray}\label{abcd}
	&a&=b_{P}e^{b_{P}\delta_{P}}[4b_{P}^{2}e^{b_{P}\delta_{P}}+\pi^{2}(e^{b_{P}\delta_{P}}-\epsilon_{P})], \\ \nonumber &b&=\pi^{2}(e^{b_{P}\delta_{P}}-\epsilon_{P})^{2}+4b_{P}^{2}e^{b_{P}\delta_{P}}(3e^{b_{P}\delta_{P}}-\epsilon_{P}),\\ \nonumber
	&c&=12b_{P}e^{b_{P}\delta_{P}}(e^{b_{P}\delta_{P}}-\epsilon_{P}),\\ \nonumber &d&=4(e^{b_{P}\delta_{P}}-\epsilon_{P})^{2}, \\ \nonumber &e&=4b_{P}^{2}e^{2b_{P}\delta_{P}}+\pi^{2}(e^{b_{P}\delta_{P}}-\epsilon_{P})^{2},\\ \nonumber
	&f&=8b_{P}e^{b_{P}\delta_{P}}(e^{b_{P}\delta_{P}}-\epsilon_{P}), \\ \nonumber &g&=d=4(e^{b_{P}\delta_{P}}-\epsilon_{P})^{2}.
	\end{eqnarray}
	
	In a similar way, the parameters $a_i$, $b_i$, $c_i$, $d_i$, $e_i$, $f_i$, $g_i$, and $k_i$ in Eq.~(\ref{Eq:abcdi}) may be related to those of Eqs.~(\ref{Reggeon1}-\ref{Eq:Otray}) (more complicated than in Eq.~(\ref{abcd})).
	
	Alternatively, the local slope $B(s,t)$ with unit pomeron intercept $\alpha(t=0)=1$ may be written as Ref. \cite{JS,Kholod}
	\begin{equation} \label{eq:B}
	B(s,t)=2\alpha'(t)[b+F(s,t)L],
	\end{equation}
	where 
	\begin{equation}\label{eq:B1}
	F(s,t)=1+\Phi'(\alpha)\Bigl[1+\frac{\pi^2\Phi(\alpha)}{4L}+\Phi(\alpha) L\Bigr]\Bigl[[1+\Phi(\alpha) L]^2+\frac{\pi^2}{4}\Phi^2(\alpha)\Bigr]^{-1}.
	\end{equation}
	Equivalently, using Eq.(\ref{GP}) with the trajectory Eq.~(\ref{Ptray}) we get for the slope  Eq.~(\ref{Eq:DPB}).
	
	Note  that $F(s,t)\rightarrow 1$ as $s\rightarrow \infty.$ Surprisingly, the function $F(s)$ decreases rapidly at small values of $s$, thus affecting the slope at small energies. It is close to $1$ at high energies where the pomeron dominates.
	
	\section*{References}
	\bibliography{mybibfile}
\end{document}